\documentclass[prc,preprint,showpacs,tightenlines,floatfix]{revtex4}
\usepackage{bm}
\usepackage{amssymb}
\usepackage{amsmath}
\usepackage{graphicx}

\begin{document}
\title{Impact of spin-orbit currents on the electroweak 
         skin of neutron-rich nuclei}
\author{C. J. Horowitz}\email{horowit@indiana.edu} 
\affiliation{Center for Exploration of Energy and Matter and
                  Department of Physics, Indiana University,
                  Bloomington, IN 47405, USA}
\author{J. Piekarewicz}\email{jpiekarewicz@fsu.edu} 
\affiliation{Department of Physics, Florida State University,
                  Tallahassee, FL 32306, USA}
\date{\today}
\begin{abstract}
 {\bf Background:} Measurements of neutron radii provide
 important constraints on the isovector sector of nuclear 
 density functionals and offer vital guidance in areas as 
 diverse as atomic parity violation, heavy-ion collisions, and 
 neutron-star structure.
 {\bf Purpose:} To assess the impact of spin-orbit currents 
 on the electromagnetic- and weak-charge radii of a variety 
 of nuclei. Special emphasis is placed on the experimentally
 accessible electroweak skin, defined as the difference 
 between weak-charge and electromagnetic-charge radii.
 {\bf Methods:} Two accurately calibrated relativistic mean
 field models are used to compute proton, neutron, charge, 
 and weak-charge radii of a variety of nuclei. 
 {\bf Results:} We find that spin-orbit contributions to the 
 electroweak skin of light neutron-rich nuclei, such as 
 ${}^{22}$O and ${}^{48}$Ca, are significant and result in a 
 substantial increase in the size of the electroweak skin 
 relative to the neutron skin.
 {\bf Conclusions:}  Given that spin-orbit contributions to
 both the charge and weak-charge radii of nuclei are often 
 as large as present or anticipated experimental error bars, 
 future calculations must incorporate spin-orbit currents in
 the calculation of electroweak form factors.
\end{abstract}
\smallskip
\maketitle

\section{Introduction}
\label{sec.introduction}

Recently the {\sl Lead Radius EXperiment} (``PREX'') at the Jefferson
Laboratory used parity violating electron scattering to probe the
weak-charge density of $^{208}$Pb\,\cite{Abrahamyan:2012gp,
Horowitz:2012tj}. In Born approximation, the parity violating
asymmetry---the fractional difference in cross sections for positive
and negative helicity electrons---is directly proportional to the weak
form factor, which is the Fourier transform of the weak-charge
density. Although for a heavy nucleus one must include the effects of
Coulomb distortions, these have been accurately
calculated\,\cite{Horowitz:1998vv, RocaMaza:2011pm}. Thus,
parity-violating electron scattering can be used with as much
confidence to measure the weak-charge form factor of the nucleus as
conventional electron scattering has been used throughout the years to
accurately map the electromagnetic charge distribution.  Many details
of a practical parity-violating experiment, along with a number of
theoretical corrections, have been discussed in
Ref.\,\cite{Horowitz:1999fk}. PREX demonstrated excellent control of
systematic errors and showed that accurate and model independent
measurements of weak-charge densities are now feasible.

In this paper we focus on the {\sl electroweak skin} of a nucleus 
which we define as 
\begin{equation}
  R_{\rm wskin} \!\equiv\! R_{\rm wk}-R_{\rm  ch} \,,
 \label{WkSkin}
\end{equation}
where $R_{\rm wk}$ and $R_{\rm  ch}$ are the root-mean-square radii of
the weak-charge and electromagnetic-charge densities, respectively,
Note that unlike the point-neutron and point-proton densities, the
weak-charge density---the source for the $Z^{0}$ weak boson---and 
the electromagnetic-charge density are physical observables. For
simplicity, we refer to $R_{\rm wskin}$ as the {\sl weak skin}. Given
that the charge radius of ${}^{208}$Pb is very 
accurately known\,\cite{Angeli:2004}, the determination of the weak
form factor of ${}^{208}$Pb by the PREX collaboration resulted in the 
following value for its weak skin\,\cite{Horowitz:2012tj}:
\begin{equation}
  R_{\rm wskin}({}^{208}{\rm Pb})=0.32\pm 0.18 ({\rm exp}) 
  \pm 0.03 ({\rm mod})\, {\rm fm} \,.
 \label{PREX_WkSkin}
\end{equation}
Note that the experimental error (``exp'') includes both statistical
and systematic errors while the small model error (``mod'') describes
the sensitivity in the extraction of $R_{\rm wk}$ due to uncertainties
in the surface thickness\,\cite{Horowitz:2012tj}.  Although the
statistical accuracy of the measurement was compromised by unforeseen
technical difficulties, a follow-up measurement (``PREX-II'') designed
to achieve the original $\!\pm0.05$\,fm goal has been
proposed and accepted\,\cite{PREXII:2012}.  Moreover, a fresh new
proposal has been made to use parity-violating electron scattering to
measure the weak form factor of ${}^{48}$Ca at a momentum transfer of
$q\!=\!0.778\,{\rm fm}^{-1}$. The {\sl Calcium Radius EXperiment}
(``C-REX'') is designed to constrain the weak-charge radius of
${}^{48}$Ca to an accuracy of $\!\pm0.03$\,fm\,\cite{CREX:2012}.  A
measurement in a smaller neutron-rich nucleus is desirable because the
form factor can be measured at a larger momentum transfer where the
parity-violating asymmetry is larger; for comparison, the weak form
factor in ${}^{208}$Pb was measured at $q\!=\!0.475\,{\rm fm}^{-1}$.
In addition, a precise measurement of the weak radius of ${}^{48}$Ca
may have a significant impact on nuclear structure as it provides
information that is independent and complementary to that found in
${}^{208}$Pb\,\cite{Piekarewicz:2012pp}. Furthermore, it is reasonable
to expect first-principle calculations of the structure of ${}^{48}$Ca
where the role of 3-neutron forces may be particularly interesting 
and important. Finally, in conjunction with PREX and PREX-II, C-REX
will provide vital guidance in areas as diverse as atomic parity
violation\,\cite{Pollock:1992mv,Sil:2005tg}, heavy-ion
collisions\,\cite{Tsang:2004zz,Chen:2004si,Steiner:2005rd,
Shetty:2007zg,Tsang:2008fd}, and neutron-star
structure\,\cite{Horowitz:2000xj,Horowitz:2001ya,Horowitz:2002mb,
Carriere:2002bx,Steiner:2004fi,Li:2005sr}.  Given the expected
accuracy of these pioneering experiments, it is critical to assess 
the role of {\sl ``sub-leading''} contributions to the weak skin
of these nuclei. Thus, it is the main goal of the present contribution 
to quantify the impact of spin-orbit currents on the electroweak skin
of a variety of nuclei.  We note that meson-exchange currents---which
are not considered in this contribution---can change the distribution
of both electromagnetic and weak charge.  For example
$\rho\pi\gamma$ and $\omega\sigma\gamma$ meson-exchange 
currents (MEC) are known to modify significantly the electromagnetic 
form factors of the deuteron\,\cite{Ito:1993au}. However, heavy 
mesons are unlikely to transport charge over large distances, so we 
do not expect significant MEC corrections to either $R_{\rm ch}$ or 
$R_{\rm wk}$ for a heavy nucleus. Yet MEC corrections could be more 
important for the case of the weak skin ($R_{\rm wk}\!-\!R_{\rm  ch}$) 
and this deserves further consideration.

The result depicted in Eq.\,(\ref{PREX_WkSkin}) represents a true
experimental milestone. It provides direct experimental evidence that
the weak-charge density in ${}^{208}$Pb extends further out than the
corresponding electromagnetic-charge density. That is, there is an
enhancement of weak charges (which are dominated by neutrons) relative
to electromagnetic charges (which are dominated by protons) near the
nuclear surface. A quantity that is closely related to the weak skin
of a heavy nucleus is the {\sl neutron skin}. In analogy to the weak
skin, the neutron skin is defined as the difference between the point 
neutron $R_{n}$ and point proton $R_{p}$ root-mean-square radii:
\begin{equation}
  R_{\rm nskin} \!\equiv\! R_{n}-R_{p} \;.
 \label{NSkin}
\end{equation}
The weak and neutron skins are closely related to each other because
the weak charge of a neutron ($\approx\!-1$) is much larger than the
weak charge of the proton ($\approx\!0.07)$ largely in the same way 
as the electromagnetic charge of the proton is much larger than that 
of the neutron. Indeed, to a good approximation the weak form factor 
of a single nucleon equals the negative of the electromagnetic form
factor of its isospin partner (see Appendix).
The structure of a nucleus is often modeled in terms of point nucleons
interacting via strong nuclear potentials or meson exchanges.
Therefore, nuclear-structure calculations provide predictions for the
neutron skin whereas it is the weak skin that is experimentally
accesible. In the present paper we calculate the difference between
$R_{\rm wskin}$ and $R_{\rm nskin}$ that originates from the internal
structure of the nucleon. The internal structure of the nucleon is
contained in a variety of well-measured single nucleon electromagnetic
and weak form factors. The electromagnetic size of the nucleon leads
to a charge radius of the nucleus $R_{\rm ch}$ that is slightly larger
than the point proton radius $R_{p}$ due (mostly) to the charge radius
of a single proton $r_{p}$ ($R_{\rm ch}^{2}\!\approx\!R_{p}^{2}+r_{p}^{2}$). 
Similarly, the weak radius $R_{\rm wk}$ is slightly larger than
$R_{n}$ (mostly) because of the weak-charge radius of a single
neutron. Note that we have reserved the use of upper-case 
$R$ to nuclear radii and of lower-case $r$ to the radii of single
nucleons.

The manuscript has been organized as follows. In
Sec.\,\ref{sec.formalism} we outline the formalism used to calculate
weak and electromagnetic charge form factors and densities. These form
factors are computed from one-body currents that include weak and
electromagnetic single-nucleon form factors. We discuss in detail the
contribution to both of the form factors from the spin-orbit
currents. For a recent reference on the impact of spin-orbit currents
on the charge density of light nuclei see Ref.\,\cite{Ong:2010gf}.
However, to our knowledge the role of spin-orbit currents on the
weak-charge density has never been studied. In Sec.\,\ref{sec.results}
we present results for proton, neutron, charge, and weak radii for a
variety of nuclei. Spin-orbit corrections are particularly large for
light nuclei and we observe a significant increase in the weak skin of
these nuclei relative to their neutron skin. We offer our conclusion 
in Sec.\,\ref{sec.conclusions}.

\section{Formalism}
\label{sec.formalism}

We start the formalism by writing the most general form of the
single-nucleon matrix elements of the electroweak current consistent
with Lorentz covariance and parity invariance. That
is\,\cite{Musolf:1993tb},
\begin{subequations}    
 \begin{eqnarray}
  &&\langle N({\bf p}',\!s')|\hat{J}^{\mu}_{\rm EM}|N({\bf p},\!s)\rangle =
       \overline{U}({\bf p}',\!s')\left[F_{1}\gamma^{\mu}+iF_{2}\sigma^{\mu\nu}
       \frac{q_{\nu}}{2M}\right]U({\bf p},\!s) \;, \\
  &&\langle N({\bf p}',\!s')|\hat{J}^{\mu}_{\rm NC}|N({\bf p},\!s)\rangle =
       \overline{U}({\bf p}',\!s')\left[\widetilde{F}_{1}\gamma^{\mu}+
       i\widetilde{F}_{2}\sigma^{\mu\nu}
       \frac{q_{\nu}}{2M}\right]U({\bf p},\!s) \;, 
 \end{eqnarray}
 \label{EWCurrentF}
\end{subequations}
where $U({\bf p},\!s)$ are on-shell nucleon spinors,
$q\!\equiv\!p'\!-\!p$ is the four momentum transfer to the nucleon,
and $F_{1,2}(\widetilde{F}_{1.2})$ are electromagnetic (weak-neutral
vector) Dirac and Pauli form factors, respectively.  Note that the
axial-vector component of the weak neutral current makes no
contribution to the elastic form factor of a nuclear ground state of
definite parity. As can be seen from the above equations, the formal
structure of the vector current is the same regardless of whether one 
considers electromagnetic or weak matrix elements or whether one 
considers protons or neutrons. Thus, for illustration purposes we
focus the derivation on the proton contribution to the charge form 
factor.

\subsection{Charge Form Factor}
\label{sec.EwFF}

Throughout this contribution we will assume the validity of the
impulse approximation, namely, that the single-nucleon form factors
defined in Eq.\,(\ref{EWCurrentF}) may be used without modification in
the nuclear medium. Moreover, we rely on an alternative description of
the single-nucleon electromagnetic current based on the electric and
magnetic Sachs form factors rather than on $F_{1,2}$. The two sets of
form factors are related as follows:
\begin{subequations}    
\begin{eqnarray}
 && G_{\rm E}(Q^{2})  \equiv F_{1}(Q^{2})-\tau F_{2}(Q^{2}) \;, \\ 
 && G_{\rm M}(Q^{2}) \equiv F_{1}(Q^{2})+F_{2}(Q^{2}) \;, 
\end{eqnarray}
\label{SachsFF}
\end{subequations}
where $Q^{2}\!=\!-q_{\mu}q^{\mu}\!>\!0$ and 
$\tau\!=Q^{2}/4M^{2}$. In turn, the electromagnetic current operator
may be written as
\begin{equation}
 \hat{J}^{\mu}_{\rm EM} = G_{\rm E}\gamma^{\mu}+
 \left(\frac{G_{\rm M}\!-\!G_{\rm E}}{1+\tau}\right)
 \left[\tau\gamma^{\mu}+i\sigma^{\mu\nu}\frac{q_{\nu}}{2M}\right]\;.
 \label{EMCurrentG}
\end{equation}
Given that we are interested in computing the elastic form factor of
spherical nuclei, only the density ({\sl i.e.,} the zeroth component
of the current) needs to be consider. Such density is given by
\begin{equation}
 \hat{J}^{0}_{\rm EM} \equiv \rho_{{}_{\rm EM}} = 
  G_{\rm E}\gamma^{0}+
  \left(\frac{G_{\rm M}\!-\!G_{\rm E}}{1+\tau}\right)
  \left[\tau\gamma^{0}+\gamma^{0}
  \frac{{\mbox{\boldmath $\gamma$}}\cdot{\bf q}}{2M}\right]\;.
 \label{EMDensityG}
\end{equation}
Thus, in the impulse approximation, the charge form factor of the
nucleus may written as follows:
\begin{equation}
 Z F_{\rm ch}(q)=G_{\rm E}(q^{2})F_{\rm V}(q)+
  \left(\frac{G_{\rm M}(q^{2})\!-\!G_{\rm E}(q^{2})}{1+\tau}\right)
  \left[\tau F_{\rm V}(q) + \frac{q}{2M} F_{\rm T}(q)\right] \,,
 \label{ChargeFF}
\end{equation}
where $Z$ is the number of protons and the charge form factor of the
nucleus has been normalized to $F_{\rm ch}(q\!=\!0)\!=\!1$. Note that
all nuclear structure information is contained in the elastic
(point-nucleon) vector and tensor form factors defined as
\begin{subequations}    
\begin{eqnarray}
 && F_{\rm V}(q) = \int \bar{\psi}({\bf r})e^{i{\bf q}\cdot{\bf r}}
      \gamma^{0}{\psi}({\bf r}) d^{3}r \;, \\ 
 && F_{\rm T}(q) = \int \bar{\psi}({\bf r})e^{i{\bf q}\cdot{\bf r}}
      \gamma^{0}{\mbox{\boldmath $\gamma$}}\cdot\hat{{\bf q}}\, 
      {\psi}({\bf r}) d^{3}r \;.
\end{eqnarray}
\label{NuclearFF0}
\end{subequations}
We now proceed to evaluate these two nuclear form factors using a
relativistic mean-field approximation.

\subsection{Relativistic Mean-Field Approximation}
\label{sec.RMF}
In a relativistic mean-field approximation, the eigenstates of the
Dirac equation corresponding to a spherically symmetric ground state
may be classified according to a generalized angular momentum
$\kappa$\,\cite{Serot:1984ey}.  That is, the single-particle solutions
of the Dirac equation may be written as
\begin{equation}
 {\cal U}_{n\kappa m}({\bf r}) = \frac {1}{r}
 \left( \begin{array}{c}
   \phantom{i}
   g_{n \kappa}(r){\cal Y}_{+\kappa m}(\hat{\bf r})  \\
  if_{n \kappa}(r){\cal Y}_{-\kappa m}(\hat{\bf r})
 \end{array}\right) \;,
\label{Uspinor}
\end{equation}
where $n$ and $m$ are the principal and magnetic quantum numbers,
respectively, and the spin-spherical harmonics are defined as follows:
\begin{equation}
 {\cal Y}_{\kappa m}(\hat{\bf r}) \equiv
 \langle{\hat{\bf r}}|l{\scriptstyle\frac{1}{2}}jm\rangle\;; \quad
 j = |\kappa|\!-\!\frac {1}{2} \;; \quad
 l = 
     \begin{cases}
           \kappa\;,     & {\rm if} \; \kappa>0; \\
        -1-\kappa\;,  & {\rm if} \; \kappa<0. 
     \end{cases}
 \label{curlyy}
\end{equation}
Note that for the phase convention adopted in Eq.\,(\ref{Uspinor}) 
(namely, the relative factor of $i$) both $g(r)$ and $f(r)$ are real 
functions in the case of real mean-field potentials. Further, the 
following spinor normalization has been used:
\begin{equation}
 \int {\cal U}^{\dagger}_{n\kappa m}({\bf r}) \,
        {\cal U}_{n\kappa m}({\bf r}) d^{3}r =
 \int_{0}^{\infty} \Big[g_{n\kappa}^{2}(r)+f_{n\kappa}^{2}(r)\Big] dr=1\;.
\end{equation}
Thus, in a mean-field approximation the vector and tensor form factors
may be expressed as a sum over occupied single-particle states. That
is,
\begin{subequations}    
\begin{eqnarray}
  &&F_{\rm V}(q) = 
      \sum_{n\kappa m}
      \int \overline{\cal U}_{n\kappa m}({\bf r})j_{{}_{0}}(qr)
      \gamma^{0}\,{\cal U}_{n\kappa m}({\bf r}) d^{3}r\,, \\ 
  &&F_{\rm T}(q) =
       \sum_{n\kappa m} 
       \int \overline{\cal U}_{n\kappa m}({\bf r}) j_{{}_{1}}(qr)\,
       i\gamma^{0}{\mbox{\boldmath $\gamma$}}\cdot\hat{{\bf r}}\, 
       {\cal U}_{n\kappa m}({\bf r}) d^{3}r\,.
\end{eqnarray}
\label{NuclearFF1}
\end{subequations}
Here $j_{{}_{l}}(qr)$ are spherical Bessel functions and the sum over
single-particle quantum numbers $\{n\kappa m\}$ is restricted to 
the occupied orbitals. Note that by virtue of the spherical
symmetry of the nuclear form factors, all explicit dependence on
$\hat{\bf q}$ has been eliminated through an {\sl ``angle-average''}
({\sl i.e.,} by integrating over $d\hat{\bf q}/4\pi$). This fact leads
to a considerable simplification and enables one to readily perform 
the integrals over the remaining solid angle. One finally obtains:
\begin{subequations}    
\begin{eqnarray}
  &&F_{\rm V}(q) = \sum_{n\kappa} (2j\!+\!1)\int_{0}^{\infty} 
      \Big[g^{2}_{n \kappa}(r)+f^{2}_{n \kappa}(r)\Big] j_{{}_{0}}(qr) dr\,, \\
  &&F_{\rm T}(q) = \sum_{n\kappa} 2(2j\!+\!1) \int_{0}^{\infty} 
       g_{n \kappa}(r)f_{n \kappa}(r)j_{{}_{1}}(qr) dr\,.    
\end{eqnarray}
\label{NuclearFF2}
\end{subequations}
Note that for a spherically symmetric ground state there is a third
independent {\sl ``scalar''} form factor that is identical to the
vector one except for a relative minus sign between upper and lower
components. Given the form of the electromagnetic current adopted in
Eq.~(\ref{EWCurrentF}), and ignoring possible off-shell ambiguities,
the scalar form factor plays no role in the present discussion.

In summary, nuclear charge and weak form factors---both normalized 
to one at $q\!=\!0$---are now obtained from  Eq.\,(\ref{ChargeFF}) by 
properly adding proton and neutron contributions. That is,
\begin{subequations}    
 \begin{eqnarray}
   Z F_{\rm ch}(q)&\!=\!&\sum_{i=p,n} \left(
   G_{\rm E}^{\,i}(q^{2})F_{\rm V}^{\,i}(q)+
   \left(\frac{G_{\rm M}^{\,i} (q^{2})\!-\!G_{\rm E}^{\,i} (q^{2})}{1+\tau}\right)
   \left[\tau F_{\rm V}^{\,i} (q) + 
   \frac{q}{2M} F_{\rm T}^{\,i}(q)\right] \right),\quad 
   \label{TotalChargeEmFF}\\
  Q_{\rm wk}F_{\rm wk}(q) &\!=\!&\sum_{i=p,n} \left(
   \widetilde{G}_{\rm E}^{\,i}(q^{2})F_{\rm V}^{\,i}(q)+
   \left(\frac{\widetilde{G}_{\rm M}^{\,i} (q^{2})
    \!-\!\widetilde{G}_{\rm E}^{\,i} (q^{2})}{1+\tau}\right)
   \left[\tau F_{\rm V}^{\,i} (q) + 
    \frac{q}{2M} F_{\rm T}^{\,i}(q)\right] \right),\quad
     \label{TotalChargeWkFF}
 \end{eqnarray}
 \label{TotalChargeFF}
\end{subequations}
where $Q_{\rm wk}$ is the weak-vector charge of the nucleus (see Appendix).

\subsection{Electromagnetic and Weak Charge Radii}
\label{sec.Radii}

The experimentally measurable electromagnetic $R_{\rm ch}$ and weak 
$R_{\rm wk}$ charge radii are obtained from the slope of the respective 
form factors at the origin. That is,
\begin{equation}    
   R_{\rm ch} = -6\frac{dF_{\rm ch}}{dq^{2}}\Big|_{q\!=\!0} 
    \quad {\rm and} \quad
   R_{\rm wk} = -6\frac{dF_{\rm wk}}{dq^{2}}\Big|_{q\!=\!0} \,.
 \label{Radii}
\end{equation}
Although in the next section we provide results in the form of tables
and figures, it is illuminating to discuss the particle content of both
$R_{\rm ch}$ and $R_{\rm wk}$. We start with the former. According to 
Eq.\,(\ref{TotalChargeEmFF}), the electromagnetic charge form factor
to first order in $q^{2}$ may be written as follows:
\begin{eqnarray}    
 Z F_{\rm ch}(q) &\!=\!& Z\left(1-\frac{q^{2}}{6}R_{p}^2\right)
 \left(1-\frac{q^{2}}{6}r_{p}^{2}\right)+
 \frac{q^{2}}{4M^{2}}\kappa_{p}Z(1+f_{{}_{\rm T}}^{p}) + \ldots 
  \nonumber \\
  &\!+\!& N\left(1-\frac{q^{2}}{6}R_{n}^2\right)
 \left(0-\frac{q^{2}}{6}r_{n}^{2}\right)+
 \frac{q^{2}}{4M^{2}}\kappa_{n}N(1+f_{{}_{\rm T}}^{n}) + \ldots
 \label{FchLowQ}
\end{eqnarray}
where $r_{p,n}^{2}$ and $\kappa_{p,n}$ are single-nucleon
electromagnetic mean-square radii and anomalous magnetic 
moments, respectively. Note that we have defined
\begin{subequations}    
 \begin{eqnarray}
   && f_{{}_{\rm T}}^{p} = \frac{4M}{3Z}\sum_{n\kappa} 
     (2j\!+\!1) \int_{0}^{\infty} 
     r g_{n \kappa}^{p}(r) f_{n \kappa}^{p}(r) dr \,, \\
   && f_{{}_{\rm T}}^{n} = \frac{4M}{3N}\sum_{n\kappa} 
     (2j\!+\!1) \int_{0}^{\infty} 
     r g_{n \kappa}^{n}(r) f_{n \kappa}^{n}(r) dr \,.
 \end{eqnarray}
 \label{FTlowQ}
\end{subequations}
Using Eq.\,(\ref{FchLowQ}) one can readily obtained an expression for
the electromagnetic charge radius of the nucleus. That is,
\begin{equation}
  R_{\rm ch}^{2} = 
  R_{p}^2 +  r_{p}^2 + \langle r_{p}^{2}\rangle_{\rm so}+
  \frac{N}{Z} \Big(r_{n}^2 + \langle r_{n}^{2}\rangle_{\rm so}\Big) \,,
 \label{Rch2}
\end{equation}
where {\sl ``spin-orbit''} contributions to the charge radius have
been defined as follows: 
\begin{subequations}    
 \begin{eqnarray}
   && \langle r_{p}^{2}\rangle_{\rm so} =
   -\frac{3\kappa_{p}}{2M^{2}}(1+f_{{}_{\rm T}}^{p}) \,,\\
  && \langle r_{n}^{2}\rangle_{\rm so} =
   -\frac{3\kappa_{n}}{2M^{2}}(1+f_{{}_{\rm T}}^{n}) \,.
 \end{eqnarray}
 \label{r2so0}
\end{subequations}
In the context of a relativistic mean-field approximation---where both
upper and lower components are self-consistently generated from the
Hartree equations---there are no further simplifications. However, one
can shed light into the nature of the spin-orbit contribution by
generating the lower component from the upper component by 
assuming a free-space relation. That is, 
\begin{equation}
  f_{n \kappa}(r) = \frac{1}{2M}\Big(\frac{d}{dr}+
  \frac{\kappa}{r}\Big)g_{n \kappa}(r) \;.
 \label{fandg}
\end{equation}
In this free-space limit an enormous simplification ensues, as the
tensor integrals given in Eq.\,(\ref{FTlowQ}) can be evaluated in
closed form. In this limit the spin-orbit contributions to the
charge radius become
\begin{subequations}    
 \begin{eqnarray}
   && \langle r_{p}^{2}\rangle_{\rm so} =
   -\frac{3\kappa_{p}}{2M^{2}}(1+f_{{}_{\rm T}}^{p}) 
    \rightarrow - \frac{2\kappa_{p}}{M^{2}Z}
    \sum_{n\kappa} {\rm sgn}(\kappa)l(l+1) \,,\\
  && \langle r_{n}^{2}\rangle_{\rm so} =
   -\frac{3\kappa_{n}}{2M^{2}}(1+f_{{}_{\rm T}}^{n}) 
    \rightarrow - \frac{2\kappa_{n}}{M^{2}N}
    \sum_{n\kappa} {\rm sgn}(\kappa)l(l+1) \,.
 \end{eqnarray}
 \label{r2so1}
\end{subequations}
The above results indicate that: (a) there is no
contribution from $s$-states and (b) there is an exact cancellation
between spin-orbit partners, which have the same $l$ but opposite
signs for $\kappa$.  Thus, in the particular case of ${}^{48}$Ca the
only spin-orbit contribution comes from the {\sl ``unpaired''}
$1{\rm f}_{7/2}$ neutron orbital. For this case we
obtain the following figure of merit:
\begin{equation}
  \langle r^{2}\rangle_{\rm so} \equiv
  \frac{N}{Z} \langle r_{n}^{2}\rangle_{\rm so} =
   \frac{6\kappa_{n}}{5M^{2}} \approx -0.101\,{\rm fm}^{2} \;.
 \label{Ca48so0}
\end{equation}
Thus, the spin-orbit contribution amounts to approximately
$-0.015$\,{\rm fm} of the total charge radius of ${}^{48}$Ca, which is
significantly larger than the quoted experimental
error\,\cite{Angeli:2004}. Note that in a self-consistent RMF
approximation the cancellation between spin-orbit partners, although
still large, will be incomplete.  However, there is an additional
cancellation between neutron and proton orbitals with the same quantum
numbers due to their almost equal but opposite anomalous magnetic
moments.  Ultimately, the spin-orbit contribution is dominated by the
unpaired orbitals.

To obtain the weak-charge radius $R_{\rm wk}$ one proceeds in an
analogous manner but now starting from Eq.\,(\ref{TotalChargeWkFF}).
Expanding the weak charge form factor to first order in $q^{2}$ one
obtains
\begin{eqnarray}    
 Q_{\rm wk} F_{\rm wk}(q) &\!=\!& Z\left(1-\frac{q^{2}}{6}R_{p}^2\right)
 g_{\rm v}^{\rm p} \left(1-\frac{q^{2}}{6}\tilde{r}_{p}^{2}\right)+
 \frac{q^{2}}{4M^{2}}\widetilde{\kappa}_{p}Z(1+f_{{}_{\rm T}}^{p}) + \ldots 
  \nonumber \\
  &\!+\!& N\left(1-\frac{q^{2}}{6}R_{n}^2\right)
 g_{\rm v}^{\rm n} \left(1-\frac{q^{2}}{6}\tilde{r}_{n}^{2}\right)+
 \frac{q^{2}}{4M^{2}}\widetilde{\kappa}_{n}N(1+f_{{}_{\rm T}}^{n}) + \ldots
 \label{FwkLowQ}
\end{eqnarray}
where $g_{\rm v}^{\rm p}\!=\!0.0721$ and 
$g_{\rm v}^{\rm n}\!=\!-0.9878$ are (radiatively-corrected)
single-nucleon weak-vector
charges\,\cite{Musolf:1993tb,Horowitz:2012tj}. 
Note that the weak-charge radii $\tilde{r}_{p,n}^{2}$ and anomalous
weak-magnetic moments $\widetilde{\kappa}_{p,n}$ may be expressed
exclusively in terms of the corresponding electromagnetic and
strange-quark quantities, as indicated in the Appendix. In this way,
the weak-charge radius of the nucleus can now be readily extracted
from Eq.\,(\ref{FwkLowQ}). That is,
\begin{equation}
  R_{\rm wk}^{2} = \frac{Z}{Q_{\rm wk}} 
  \Big[g_{\rm v}^{\rm p}\left(R_{p}^2 + \tilde{r}_{p}^2\right) + 
  \langle \tilde{r}_{p}^{2}\rangle_{\rm so}\Big] +
  \frac{N}{Q_{\rm wk}} 
  \Big[g_{\rm v}^{\rm n}\left(R_{n}^2 + \tilde{r}_{n}^2\right) + 
  \langle \tilde{r}_{n}^{2}\rangle_{\rm so}\Big]
 \label{Rwk1}
\end{equation}
where spin-orbit contributions to the weak-charge radius 
are defined in analogy to Eq.\,(\ref{r2so1}): 
\begin{subequations}    
 \begin{eqnarray}
   && \langle\tilde{r}_{p}^{2}\rangle_{\rm so} =
   -\frac{3\widetilde{\kappa}_{p}}{2M^{2}}(1+f_{{}_{\rm T}}^{p}) \,,\\
  && \langle \tilde{r}_{n}^{2}\rangle_{\rm so} =
   -\frac{3\widetilde{\kappa}_{n}}{2M^{2}}(1+f_{{}_{\rm T}}^{n}) \,.
 \end{eqnarray}
 \label{r2sowk}
\end{subequations}
Given that 
$Q_{\rm wk}\!=\!Z g_{\rm v}^{\rm p}\!+\!N g_{\rm v}^{\rm n}$ 
(with $g_{\rm v}^{\rm p}\!\approx\!0$ and 
$g_{\rm v}^{\rm n}\!\approx\!-1$) the weak-charge radius 
of the nucleus $R_{\rm wk}$ is dominated by the contribution 
from the neutron radius $R_{n}$.  

We close this section by providing a figure of merit for the
weak-charge radius of ${}^{48}$Ca by neglecting the strange-quark
contribution and by assuming the free-space relation in the evaluation
of the tensor integrals [Eq.\,(\ref{FTlowQ})]. We obtain:
\begin{equation}
  R_{\rm wk}^{2} = \frac{Z}{Q_{\rm wk}} 
  \Big[g_{\rm v}^{\rm p}R_{p}^2 + 
         g_{\rm v}^{\rm p}r_{p}^2 +
         g_{\rm v}^{\rm n}r_{n}^2\Big] +
  \frac{N}{Q_{\rm wk}} 
  \Big[g_{\rm v}^{\rm n}R_{n}^2 + 
         g_{\rm v}^{\rm n}r_{p}^2 +
         g_{\rm v}^{\rm p}r_{n}^2 +
  \frac{6}{7M^{2}}(g_{\rm v}^{\rm n}\kappa_{p}+
                          g_{\rm v}^{\rm p}\kappa_{n})\Big]\,.
 \label{Rwk1}
\end{equation}
Note that in the above expression single-nucleon mean-square radii
($r_{p,n}^{2}$) as well as anomalous magnetic moments ($\kappa_{p,n}$)
are purely electromagnetic. Assuming (point) proton and neutron radii
for ${}^{48}$Ca as predicted by the FSUGold
model\,\cite{Todd-Rutel:2005fa} (or ``FSU'' for short) we obtain
\begin{equation}
    R_{\rm wk} \!=\!3.679 (3.669)\,{\rm fm} \;,
 \label{RwkCa48}
\end{equation}
where the quantity in parenthesis represents the FSU prediction 
without including the spin-orbit contribution that amounts to:
\begin{equation}
  \langle \tilde{r}^{2}\rangle_{\rm so} \equiv
  \frac{N}{Q_{\rm wk}} \langle \tilde{r}_{n}^{2}\rangle_{\rm so} =
    \frac{N}{Q_{\rm wk}} 
    \frac{6(g_{\rm v}^{\rm n}\kappa_{p}+
                g_{\rm v}^{\rm p}\kappa_{n})}{7M^{2}}
  \approx +0.077\,{\rm fm}^{2} \;.
 \label{Ca48so2}
\end{equation}
Given that the weak spin-orbit contribution has the opposite sign as
the corresponding electromagnetic one [see Eq.\,(\ref{Ca48so0})], the
spin-orbit contribution to the weak skin of ${}^{48}$Ca is
significant; of about 0.03\,fm (see Table\,\ref{Table2}). This is
within the projected accuracy of the proposed C-REX
experiment\,\cite{CREX:2012}.

\subsection{Electroweak Nucleon Form Factors}
\label{sec.NucleonFF}

Since our main objective is the calculation of electromagnetic- and
weak-charge radii, we adopt a simple dipole parametrization of the
single nucleon form factors that is accurate at moderate values of the 
momentum transfer. That is, we define electromagnetic single nucleon
form factors as follows\,\cite{Kelly:2002if}:
\begin{equation}
   G_{\rm E}^{\,p}(Q^{2}) = 
   \frac{G_{\rm M}^{\,p}(Q^{2})}{\mu_{p}} = 
   \frac{G_{\rm M}^{\,n}(Q^{2})}{\mu_{n}} = G_{\rm D}(Q^{2}) \;,
\label{NucleonFF}
\end{equation}
where the dipole form factor is given by
\begin{equation}
    G_{\rm D}(Q^{2}) = \left(1+\frac{Q^{2}}{12}r_{p}^{2}\right)^{-2} \;.
 \label{GDipole}
\end{equation}
Here $r_{p}^{2}\!=\!0.769\,{\rm fm}^{2}$ is the mean-square proton
radius, $\mu_{p}\!=\!2.793$ the proton magnetic moment, and
$\mu_{n}\!=\!-1.913$ the neutron magnetic moment. For the
electromagnetic neutron form factor---which vanishes at
$Q^{2}\!=\!0$---we rely on the following Galster
parametrization\,\cite{Kelly:2002if}:
\begin{equation}
   G_{\rm E}^{\,n}(Q^{2}) = -\left(\frac{Q^{2}r_{n}^{2}/6}{1+Q^{2}/M^{2}}\right)
  G_{\rm D}(Q^{2})  \;,
\label{NeutronFF}
\end{equation}
where $r_{n}^{2}\!=\!-0.116\,{\rm fm}^{2}$ is the electromagnetic
mean-square radius of the neutron. Note that all form factors are
expressed exclusively in terms of experimentally determined single
nucleon mean-square radii and magnetic
moments\,\cite{Nakamura:2010zzi}. Finally, given that the
strange-quark form factors of the nucleon are small at small momentum
transfers, we will ignore them in this contribution.  In this case,
the weak-charge form factors of the nucleon can be expressed
exclusively in terms of the corresponding electromagnetic form factors
given here and the weak charges of the nucleon (see Appendix). For
example, using Eqs.(\ref{Weakr2s}) and\,(\ref{Weakmus}) the weak 
mean-square radii and magnetic moments are given by 
\begin{subequations}    
 \begin{eqnarray}
   && \tilde{r}_{p}^{2}=2.358\,{\rm fm}^{2}\,,  \quad  \tilde{\mu}_{p}=+2.091\,;\\
   && \tilde{r}_{n}^{2}=0.777\,{\rm fm}^{2}\,,  \quad  \tilde{\mu}_{n}=-2.897\,.
\end{eqnarray}
 \label{WeakNFF}
\end{subequations}

\section{Results}
\label{sec.results}

In this section we present results---primarily proton, neutron,
charge, and weak-charge radii---for a variety of nuclei as predicted
by two accurately calibrated relativistic mean field models:
FSU\,\cite{Todd-Rutel:2005fa} and
NL3\,\cite{Lalazissis:1996rd,Lalazissis:1999}.
We start by displaying in Table\,\ref{Table1} the contribution from
the individual single-particle orbitals in ${}^{48}$Ca to the charge
and weak-charge spin-orbit radius. For illustration purposes, the
predictions have been made using only the FSU parametrization.  The
quantities in parenthesis represent the results obtained by assuming a
free-space relation between upper and lower components, as indicated
in Eq.\,(\ref{fandg}). Qualitatively, all major trends in the results
may be understood using this simplified case.  In particular, the
results displayed in parenthesis in Table\,\ref{Table1}: (i) are
independent of the dynamics, (ii) scale as $l(l\!+\!1)$, and (iii)
display an exact cancellation among spin-orbit partners [see
Eqs.\,(\ref{r2so1})]. Thus, in this limit the sole contribution to the
spin-orbit radius comes from unpaired spin-orbit partners (an unpaired
$1{\rm f}_{7/2}$ neutron orbital in the case of ${}^{48}$Ca). Given
the natural ordering of spin-orbit partners in nuclei, namely, the
$\kappa\!<\!0$ orbital more deeply bound than the $\kappa\!>\!0$
orbital, the spin-orbit contribution to the charge radius is always
negative in the case of unpaired neutrons. Note that the situation is
reversed in the case of protons due to an anomalous magnetic moment of
opposite sign to that of the neutron. Moreover, in the case of the
weak-charge radius, the spin-orbit contribution is always of opposite
sign---for both neutrons and protons---than in the electromagnetic
case. Indeed, for a given nucleon orbital the ratio of spin-orbit
contributions is given by the following simple expression:
\begin{equation}
  \frac{\langle \tilde{r}_{p,n}^{2}\rangle_{\rm so}} 
   {\langle {r}_{p,n}^{2}\rangle_{\rm so}} =
   \left(\frac{\widetilde{\kappa}_{p,n}}{\kappa_{p,n}}\right)
   \left(\frac{Z}{Q_{\rm wk}}\right) \approx
   \left(\frac{\kappa_{n,p}}{\kappa_{p,n}}\right) 
   \left(\frac{Z}{N}\right) <0 \;.
 \label{WkEMratio}
\end{equation}
Hence, although in general small, charge and weak-charge 
spin-orbit radii contribute with the same sign to the weak 
skin of a nucleus. For example, as indicated in 
Table\,\ref{Table2} the spin-orbit contribution amounts to
about 0.04\,fm in the case of the weak skin of ${}^{22}$O and 
to about $0.03\,$fm for ${}^{48}$Ca.

\begin{widetext}
\begin{center}
\begin{table}[h]
\begin{tabular}{|c||r|r|r|}
 \hline
 $nlj\,(\kappa)$ &$\langle r^{2}_{p}\rangle_{\rm so}\hfil$ 
             &$\langle r_{n}^{2}\rangle_{\rm so}\hfil$
             &$\langle r^{2}\rangle_{\rm so}\hfil$ \\
 \hline
 \hline
 $1{\rm s}_{1/2}\,(-1)$ &$+0.570(+0.000)$
                          &$-0.648(+0.000)$
                          &$-0.078(+0.000)$ \\
 $1{\rm p}_{3/2}\,(-2)$ &$+3.107(+1.583)$
                          &$-3.409(-1.690)$
                          &$-0.302(-0.107)$\\
 $1{\rm p}_{1/2}\,(+1)$ &$-1.997(-1.583)$
                          &$+2.113(+1.690)$
                          &$+0.116(+0.107)$\\
 $1{\rm d}_{5/2}\,(-3)$ &$+7.138(+4.750)$	
                          &$-7.710(-5.069)$
	                  &$-0.572(-0.319)$\\
 $1{\rm d}_{3/2}\,(+2)$ &$-6.205(-4.750)$	
                          &$+6.613(+5.069)$	
                          &$+0.408(+0.319)$\\
 $2{\rm s}_{1/2}\,(-1)$ &$+0.178(+0.000)$	
                          &$-0.177(+0.000)$	
                          &$+0.001(+0.000)$\\
 $1{\rm f}_{7/2}\,(-4)$ &$+0.000(+0.000)$	
                         &$-13.042(-10.138)$	
                         &$-13.042(-10.138)$\\
\hline
 Total                &$+2.791(+0.000)$	
                         &$-16.260(-10.138)$	
                         &$-13.469(-10.138)$\\
\hline
\hline
 $nlj\,(\kappa)$ &$\langle \tilde{r}^{2}_{p}\rangle_{\rm so}\hfil$ 
             &$\langle \tilde{r}_{n}^{2}\rangle_{\rm so}\hfil$
             &$\langle \tilde{r}^{2}\rangle_{\rm so}\hfil$ \\
 \hline
 \hline
 $1{\rm s}_{1/2}\,(-1)$ &$-0.489(+0.000)$
                          &$+0.494(+0.000)$
                          &$+0.005(+0.000)$\\
 $1{\rm p}_{3/2}\,(-2)$ &$-2.669(-1.360)$
                          &$+2.595(+1.286)$
                          &$-0.074(-0.074)$\\
 $1{\rm p}_{1/2}\,(+1)$ &$+1.716(+1.360)$
                          &$-1.608(-1.286)$
                          &$+0.108(+0.074)$\\
 $1{\rm d}_{5/2}\,(-3)$ &$-6.132(-4.081)$	
                          &$+5.869(+3.859)$
	                  &$-0.263(-0.222)$\\
 $1{\rm d}_{3/2}\,(+2)$ &$+5.331(+4.081)$	
                          &$-5.034(-3.859)$	
                          &$+0.297(+0.222)$\\
 $2{\rm s}_{1/2}\,(-1)$ &$-0.153(+0.000)$	
                          &$+0.135(+0.000)$	
                          &$-0.018(+0.000)$\\
 $1{\rm f}_{7/2}\,(-4)$ &$+0.000(+0.000)$	
                         &$+9.928(+7.717)$	
                         &$+9.928(+7.717)$\\
\hline
 Total                &$-2.396(+0.000)$	
                         &$+12.379(+7.717)$	
                         &$+9.983(+7.717)$\\
\hline
\end{tabular}
\caption{Contributions from the individual single-particle 
orbitals to the spin-orbit component of the mean square 
charge radius (upper table) and weak-charge radius (lower
table) of ${}^{48}$Ca. Mean square radii are expressed 
in units of $10^{-2}\,{\rm fm}^{2}$ and were generated using 
the FSU interaction. Quantities displayed in parenthesis have 
been computed using the free-space relation given in
Eq.\,(\ref{fandg}).}
\label{Table1}
\end{table}
\end{center}
\end{widetext}

Also shown in Table\,\ref{Table1} are predictions from the
self-consistent RMF approach, where now both upper and lower
components are dynamically generated. The qualitative trends 
discussed previously are clearly preserved, although now the 
cancellation among spin-orbit partners is incomplete.  For 
example, the contributions from the $1{\rm p}$ orbitals 
differ from each other by about 55-60\%. However, although 
the spin-orbit cancellation is incomplete, there is an additional 
cancellation stemming from the nucleon anomalous magnetic 
moments (which are nearly equal in magnitude but opposite in 
sign). Ultimately, the spin-orbit radius continues to be dominated 
by the unpaired $1{\rm f}_{7/2}$ neutron orbital.  Note, however, 
that the exact RMF prediction exceeds by about 30\% the analytic 
result obtained from assuming the free-space relation. This
enhancement is due to the reduction of the effective nucleon 
mass in the nuclear medium.

\begin{widetext}
\begin{center}
\begin{table}[h]
\begin{tabular}{|c||c||c|c|c|c|c|c|}
 \hline
 Nucleus & Model  & $R_{p}$  & $R_{n}$ & $R_{\rm ch}$ & $R_{\rm wk}$ 
               & $R_{\rm nskin}$  & $R_{\rm wskin}$ \\     
 \hline
 \hline
${}^{22}$O   &NL3&2.593&3.026&2.671(2.700)&3.172(3.158)&0.433&0.502(0.458)\\
                   &FSU&2.580&2.997&2.658(2.688)&3.144(3.129)&0.417&0.487(0.442)\\	
\hline
 ${}^{48}$Ca&NL3&3.379&3.605&3.449(3.467)&3.724(3.711)&0.226&0.275(0.243)\\
                   &FSU&3.366&3.563&3.435(3.455)&3.683(3.669)&0.197&0.247(0.214)\\
\hline
 ${}^{90}$Zr &NL3&4.194&4.308&4.254(4.268)&4.404(4.393)&0.114&0.149(0.126)\\
                  &FSU&4.181&4.269&4.242(4.255)&4.364(4.353)&0.088&0.123(0.098)\\
\hline 
${}^{118}$Sn&NL3&4.561&4.760&4.636(4.628)&4.835(4.843)&0.199&0.199(0.215)\\
                  &FSU&4.559&4.707&4.634(4.625)&4.780(4.788)&0.148&0.147(0.162)\\
\hline
${}^{132}$Sn&NL3&4.643&4.989&4.700(4.705)&5.077(5.074)&0.346&0.377(0.369)\\
                  &FSU&4.654&4.925&4.710(4.716)&5.011(5.008)&0.271&0.301(0.292)\\
\hline
${}^{208}$Pb&NL3&5.460&5.740&5.510(5.514)&5.815(5.814)&0.280&0.305(0.300)\\
                  &FSU&5.469&5.676&5.519(5.523)&5.749(5.747)&0.207&0.230(0.224)\\
\hline
${}^{138}$Ba&NL3&4.776&5.012&4.827(4.838)&5.100(5.093)&0.237&0.273(0.255)\\
                  &FSU&4.775&4.957&4.826(4.837)&5.043(5.035)&0.182&0.217(0.198)\\
\hline
${}^{158}$Dy&NL3&5.039&5.235&5.099(5.098)&5.309(5.311)&0.196&0.210(0.212)\\
                  &FSU&5.027&5.172&5.087(5.086)&5.245(5.246)&0.146&0.158(0.160)\\
\hline
${}^{176}$Yb&NL3&5.215&5.497&5.273(5.272)&5.573(5.574)&0.282&0.300(0.302)\\
                  &FSU&5.208&5.424&5.266(5.265)&5.498(5.498)&0.215&0.232(0.233)\\
\hline
\end{tabular}
\caption{Proton, neutron, charge, and weak-charge radii (in fm) 
 of a variety of nuclei as predicted by the NL3 and FSU relativistic 
 mean field models. The last two columns display the neutron skin 
 ($R_{n}$-$R_{p}$) and weak skin ($R_{\rm wk}$-$R_{\rm  ch}$),
 respectively. Quantities displayed in parenthesis have been 
 computed without the spin-orbit correction.}
\label{Table2}
\end{table}
\end{center}
\end{widetext}

In Table\,\ref{Table2} predictions are displayed for the
root-mean-square radii for a variety of magic (or semi-magic) nuclei
as predicted by the NL3 and
FSU models. Also shown (last three entries) are nuclei of relevance to
the atomic-parity violating program~\cite{Sil:2005tg}.  These nuclei
are relevant as they are members of long chains of naturally occurring
isotopes that help eliminate uncertainties in the atomic theory by
forming suitable ratios of parity-violating observables.  By doing so,
uncertainties in neutron radii become the limiting factor in the
search of physics beyond the standard model.  Note that the NL3 and
FSU models---although both accurately calibrated---predict large
differences in neutron radii. However, we caution that in the case of
${}^{158}$Dy and ${}^{176}$Yb, nuclear deformation and pairing
correlations---which have not been included---may play an important
role.

The first three columns of numbers in Table\,\ref{Table2} provide
predictions for the point-proton, point-neutron, and charge radii of
several nuclei. To a large extent these results are expected and in
some particular cases (such as ${}^{208}$Pb) have been extensively
discussed. For example, given that the charge form factor of many of
these nuclei has been accurately measured via (parity conserving)
elastic electron scattering, this information---mostly in the form of
charge radii---has been incorporated into the calibration of the RMF
models. Thus, the predictions of both models for charge radii are in
fairly good agreement ({\sl e.g.,} they differ by only 0.009\,fm in
the case of ${}^{208}$Pb).  In contrast, the lack of reliable neutron
form factors leaves the isovector sector of the relativistic
functionals largely unconstrained, thereby generating large
differences in the predictions of neutron radii ({\sl e.g.,}
0.064\,fm, or more than 1\%, in the case of ${}^{208}$Pb).
Fortunately, the prospects of constraining the isovector sector
through the measurement of neutron-radii are very good. Indeed, the
vigorous and highly successful parity-violating program at JLAB has
recently provided the first model-independence evidence in favor of a
weak skin in ${}^{208}$Pb. Moreover, a follow-up measurement has been
approved to achieve the original $\!\pm0.05$\,fm goal (PREX-II) and a
fresh new one (C-REX) has been proposed to constrain the neutron
radius of ${}^{48}$Ca to $\!\pm0.03$\,fm. It is therefore critical and
timely to assess the impact of the spin-orbit corrections on the weak
radius of these nuclei---which until now has never been considered.

\begin{figure}[ht]
\vspace{-0.05in}
\includegraphics[width=0.49\columnwidth,angle=0]{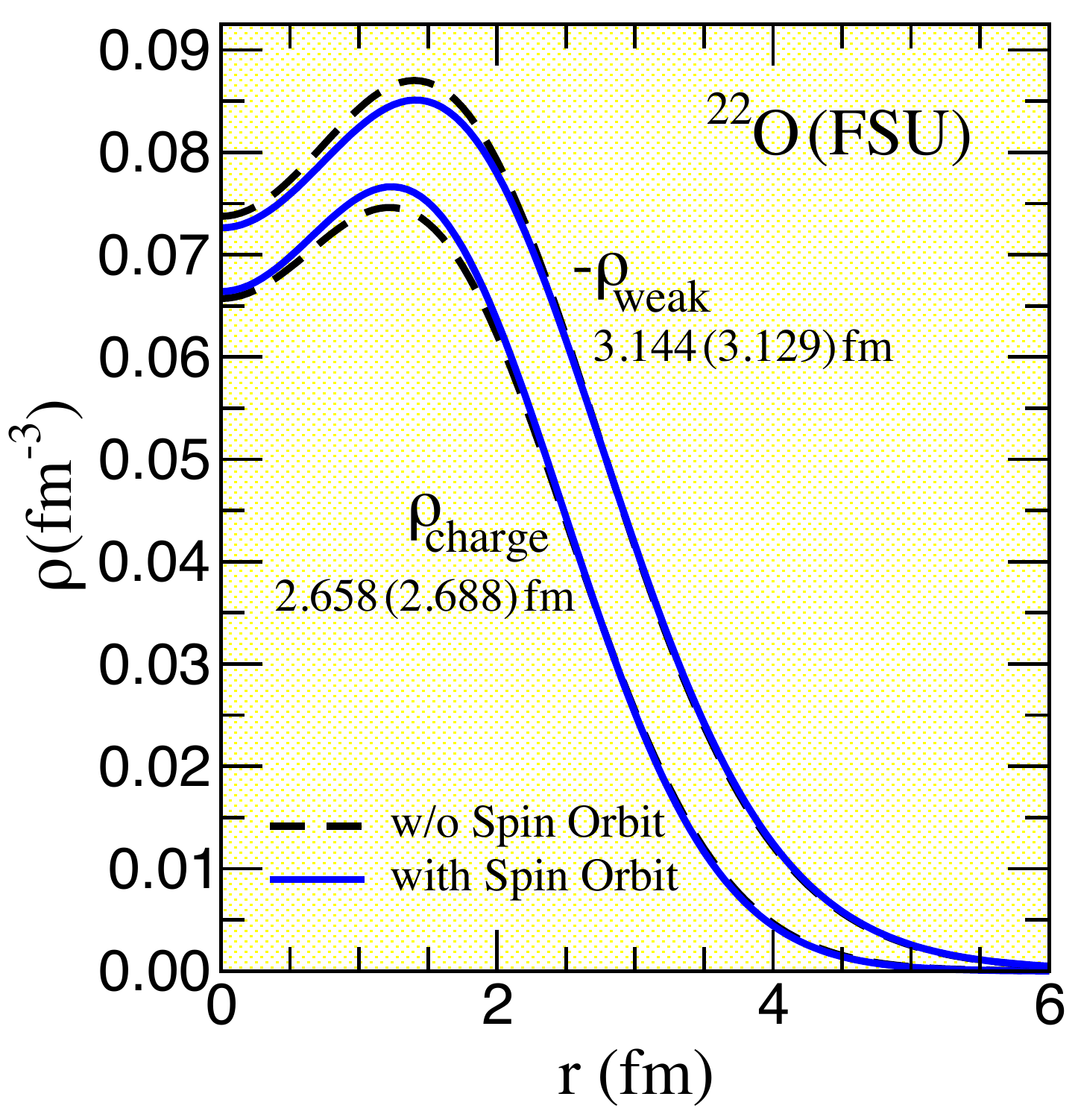}
\includegraphics[width=0.49\columnwidth,angle=0]{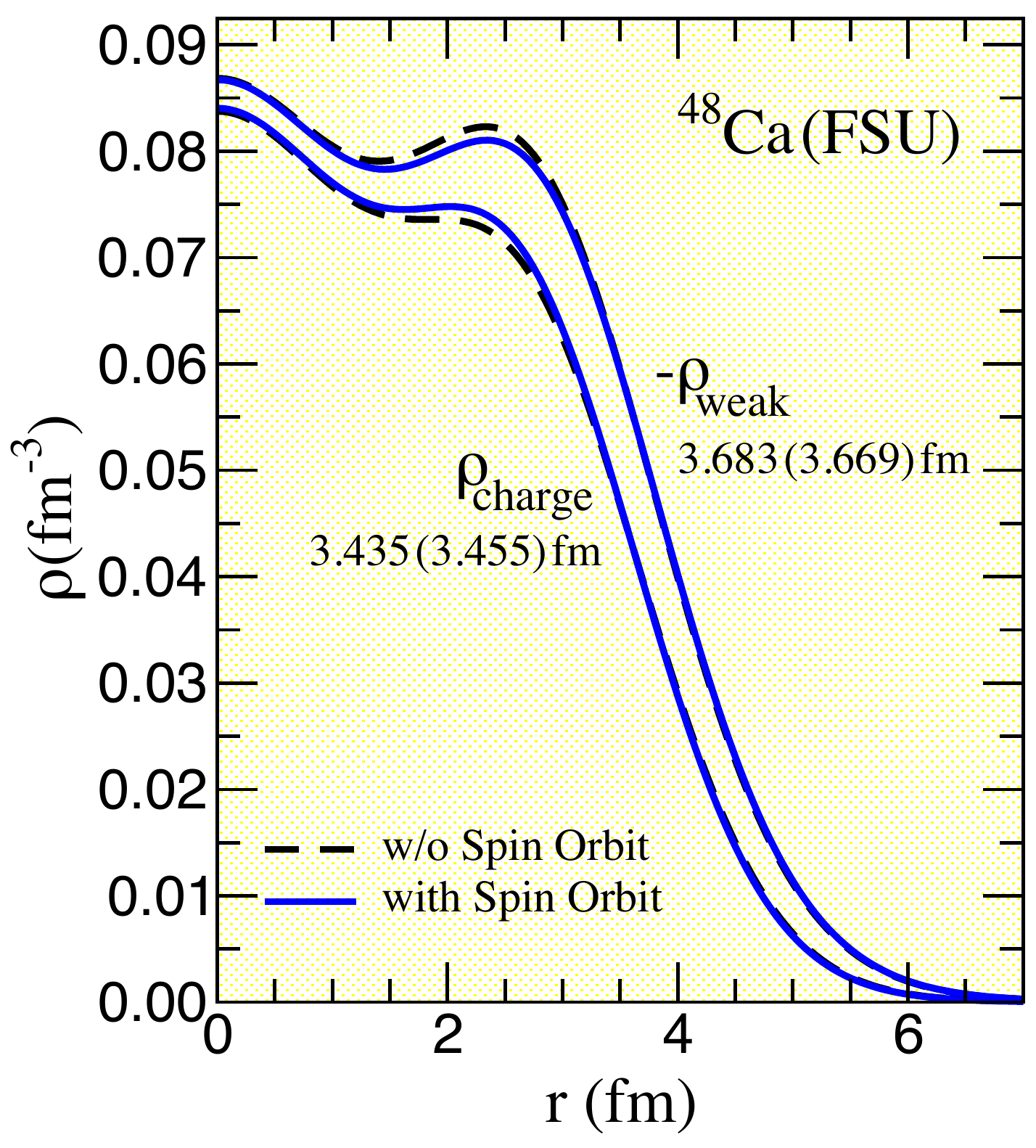}
\caption{(Color online). Charge and weak-charge densities for
${}^{22}$O and ${}^{48}$Ca as predicted by the FSU interaction.
The dashed lines represent the corresponding densities in the
absence of spin-orbit corrections.}
\label{Fig1}
\end{figure}

\begin{figure}[ht]
\vspace{-0.05in}
\includegraphics[width=0.49\columnwidth,angle=0]{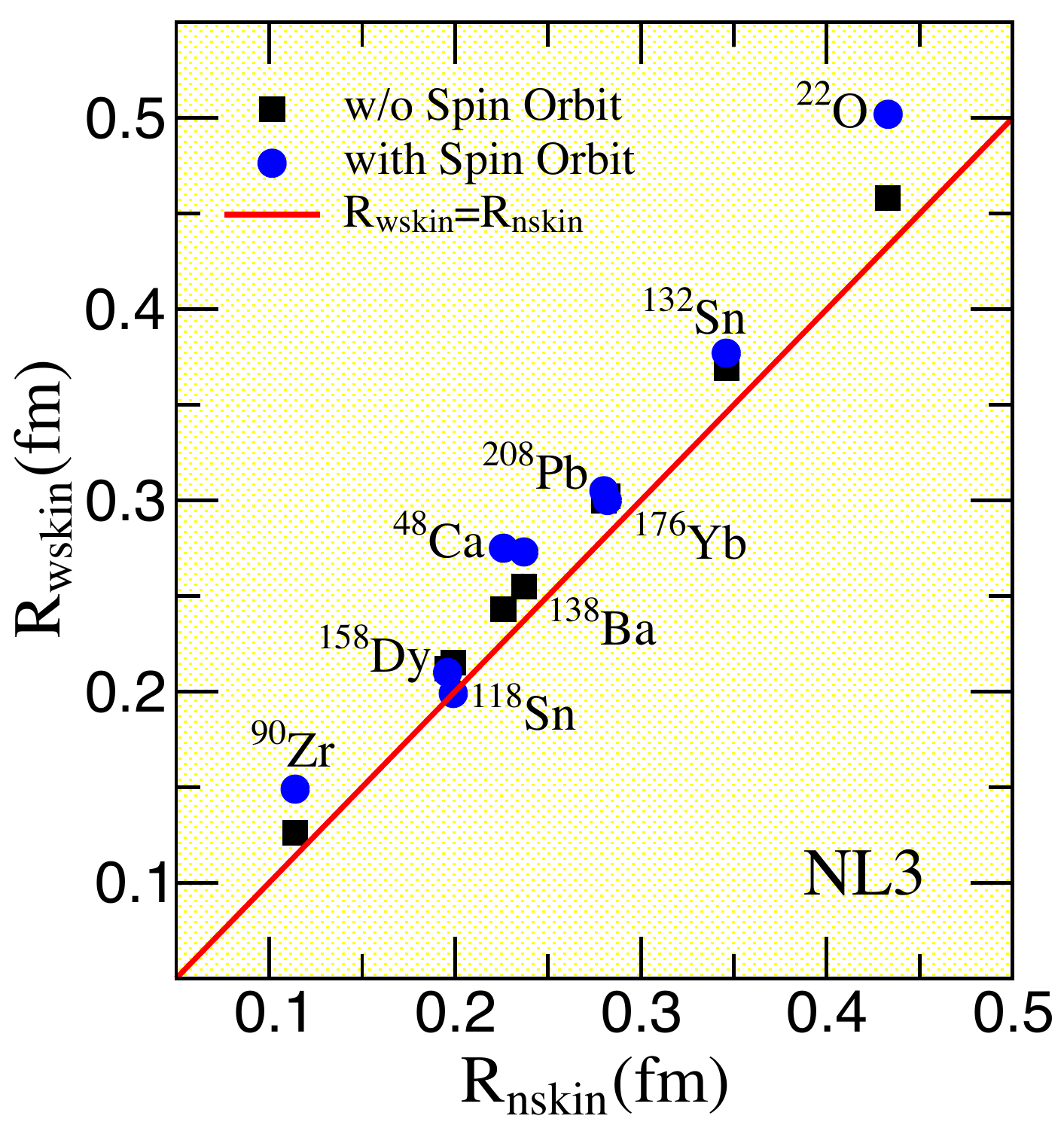}
\includegraphics[width=0.49\columnwidth,angle=0]{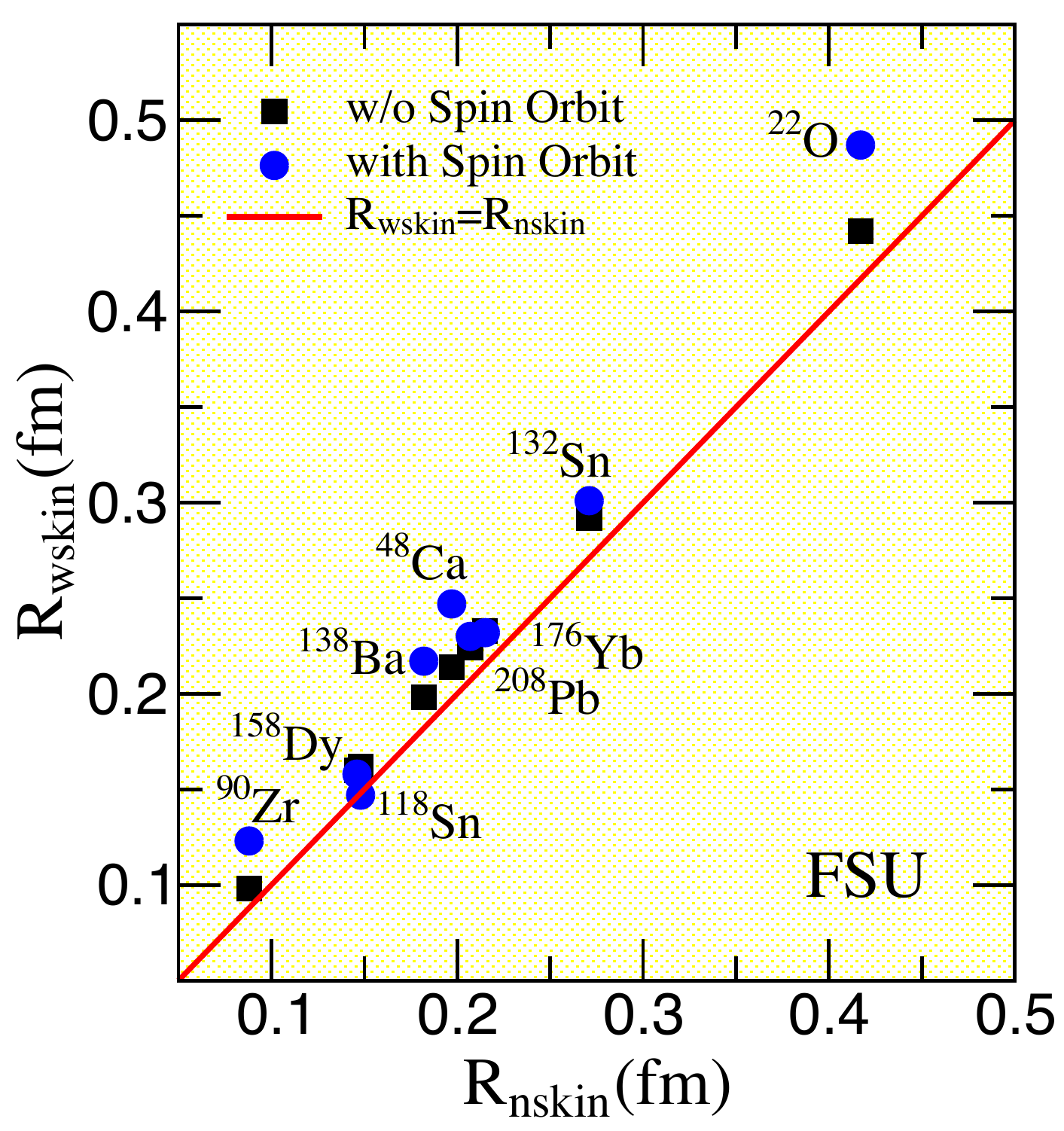}
\caption{(Color online). Electroweak skin ($R_{\rm wk}$-$R_{\rm  ch}$)
with and without spin-orbit corrections as a function of neutron 
skin ($R_{n}$-$R_{p}$) for the various neutron-rich nuclei considered 
in this work. Predictions are made using both the NL3 and FSU 
interactions.}
\label{Fig2}
\end{figure}

Given the enormous cancellation among spin-orbit partners and
proton-neutron orbitals with the same quantum numbers, the 
spin-orbit contribution to the charge and weak-charge radii lacks 
the coherence displayed by the dominant vector form factor [see
Eqs.\,(\ref{NuclearFF2})]. Thus, one expects that the spin-orbit
correction will be largest for light nuclei with unpaired spin-orbit
partners. Indeed, Table\,\ref{Table2} displays relatively large
spin-orbit contributions---of $\approx 0.045\,$fm and $\approx
0.03\,$fm---to the weak skin of ${}^{22}$O and ${}^{48}$Ca,
respectively. These spin-orbit corrections are commensurate with the
projected experimental uncertainty so they must be included in the
prediction of both charge and weak-charge radii. Charge and
weak-charge densities for ${}^{22}$O and ${}^{48}$Ca---with and
without spin-orbit corrections---as predicted by the FSU interaction
are also shown in Fig.\,\ref{Fig1}.  Note that the effect from
neglecting the spin-orbit contribution is clearly
discernible in the figure, as is the fact that the modification to the
charge and weak-charge radii goes in the opposite direction---thereby
enhancing its contribution to the weak skin.  As alluded earlier, the
impact of the spin-orbit contribution diminishes with increasing
baryon number and amounts to only $\approx\!0.002\,$fm in the case of
the weak radius of ${}^{208}$Pb---significantly below the anticipated
$\!\pm0.05$\,fm of the PREX-II measurement. Yet, the
$\approx\!0.004\,$fm contribution to the charge radius of ${}^{208}$Pb
is significantly larger than the minute $0.0009$\,fm error quoted in
Ref.\,\cite{Angeli:2004}. In the particular case of atomic nuclei of
relevance to the atomic parity violating program, the spin-orbit
contribution to the weak-charge radius of ${}^{138}$Ba, ${}^{158}$Dy,
and ${}^{176}$Yb amounts to $\approx\!0.008\,$fm,
$\approx\!-0.002\,$fm, and $\approx\!-0.001$\,fm,
respectively. Note that whereas the NL3 and FSU predictions
for the weak skin of the various nuclei differ significantly, the
spin-orbit contribution appears to have very little model dependence.
In general, we find that the intrinsic structure of the nucleon leads
to a weak skin that is larger than the corresponding neutron skin. 
Moreover, for light neutron-rich nuclei---such as ${}^{22}$O and
${}^{48}$Ca---the spin-orbit contribution generates a further
enhancement of the weak skin. In particular, whereas the FSU
interaction predicts a neutron skin in ${}^{22}$O of 
$R_{\rm nskin}\!=\!0.417\,$fm, the prediction for the experimentally 
measurable weak skin is significantly larger, namely, 
$R_{\rm wskin}\!=\!0.487\,$fm. Although not as large, the effect
is still significant in ${}^{48}$Ca ($R_{\rm nskin}\!=\!0.197\,$fm 
and $R_{\rm wskin}\!=\!0.247\,$fm) and even in ${}^{208}$Pb 
($R_{\rm nskin}\!=\!0.207\,$fm and $R_{\rm wskin}\!=\!0.230\,$fm).
We also display some of these results in graphical form in
Fig.\,\ref{Fig2} where predictions for the weak skin---with and
without spin-orbit contributions---are shown as a function of the
corresponding neutron skin. Again, the neutron skin is
interesting as it represents a pristine nuclear-structure observables
that is instrumental in constraining the isovector sector of the
nuclear density functional. The weak skin, on the other hand, although
sensitive to the internal structure of the nucleon, is both
experimentally accessible as well as strongly correlated to the
neutron skin. Note that in Fig.\,\ref{Fig2} we have added the line
$R_{\rm wskin}\!=\!R_{\rm nskin}$ to indicate that in all cases
(except for ${}^{118}$Sn) the weak skin is larger than the neutron
skin---and significantly larger for the lighter neutron-rich
nuclei. Finally, the fact that the points are more ``compressed'' in the
case of FSU than NL3 is a reflection of the softer symmetry energy of
the former relative to the latter.

\section{Conclusions}
\label{sec.conclusions}
 
A relativistic mean-field approximation has been used to compute
proton, neutron, charge, and weak-charge densities and form factors of
a variety of neutron-rich nuclei. Special emphasis has been placed on
the impact of spin-orbit currents on the electroweak skin of these
nuclei. Although closely related to the neutron skin, the weak
skin---defined as the difference between the experimentally accesible
weak- and electromagnetic-charge radii---is sensitive to the internal
structure of the nucleon.  The weak-charge radius of a nucleus is
closely related to the (point) neutron radius because the weak charge
of the neutron is much larger than that of the proton.  This is
analogous to the reason why the charge and proton radii are closely
related. In the absence of spin-orbit corrections, and regardless of 
the model, we found a weak skin larger than the neutron skin for all
nuclei investigated here.

Once spin-orbit currents were incorporated, a significant increase in
the weak skin---especially in the case of the two lighter nuclei
${}^{22}$O and ${}^{48}$Ca---was observed. In particular, we found
quite generally that the spin-orbit contribution to the weak and
charge radii enter with opposite sign so their impact is enhanced when
computing the weak skin. For example, in the case of ${}^{48}$Ca the
spin-orbit enhancement amounts to about 0.03\,fm; this is commensurate
with the projected accuracy of the C-REX experiment.  The impact of
the spin-orbit contribution decreases with increasing mass number, so
its effect in ${}^{208}$Pb is small (0.006\,fm) and well below the
proposed $\pm 0.05\,$fm accuracy.  The spin-orbit contribution is in
general small due to the strong cancellation between spin-orbit
partners and between neutron-proton pairs in identical single-particle
orbits. Given that it lacks the coherence displayed by the dominant
(vector) contribution, spin-orbit currents have the largest impact on
light neutron-rich (or proton-rich) nuclei with unpaired spin-orbit
partners, as in the case of the $1{\rm f}_{7/2}$ neutron orbital in
${}^{48}$Ca.  We note that in the RMF case where the cancellation
among spin-orbit partners is incomplete, there is a further
enhancement (of about 30\%) relative to the non-relativistic
predictions due to the presence of a strongly attractive scalar
potential (the so-called ``$M^{\star}$-effect'').

In summary, accurately calibrated relativistic mean field models 
have been used to compute the impact of spin-orbit currents on 
the electroweak skin of a variety of nuclei.  Given that spin-orbit
contributions to both the charge and weak-charge radii may be
significant and often as large as existent or projected experimental 
uncertainties, spin-orbit currents should be routinely incorporated
into future calculations of both charge and weak-charge form factors.

\begin{acknowledgments}
  This work was supported in part by grants from the U.S. Department
  of Energy DE-FG02-87ER40365 and DE-FD05-92ER40750.
\end{acknowledgments}

\appendix*
\section{Single-nucleon electroweak currents}
In terms of the underlying quark vector currents, the electromagnetic
and weak neutral currents displayed in Eq.\,(\ref{EWCurrentF}) are given
by the following expressions\,\cite{Musolf:1993tb}:
\begin{subequations}    
 \begin{eqnarray}
  && \hat{J}^{\mu}_{\rm EM} = 
  \sum_{\rm f=1}^{3} Q_{\rm f}\bar{q}_{\rm f}\gamma^{\mu}q_{\rm f} =
  \frac{2}{3}\bar{u}\gamma^{\mu}u -
  \frac{1}{3}\bar{d}\gamma^{\mu}d -
  \frac{1}{3}\bar{s}\gamma^{\mu}s 
  \label{JEM}\,,\\
  && \hat{J}^{\mu}_{\rm NC} = 
  \sum_{\rm f=1}^{3} g_{\rm v}^{\rm f}\bar{q}_{\rm f}\gamma^{\mu}q_{\rm f} =
  g_{\rm v}^{\rm u}\bar{u}\gamma^{\mu}u +
  g_{\rm v}^{\rm d}\bar{d}\gamma^{\mu}d +
  g_{\rm v}^{\rm s}\bar{s}\gamma^{\mu}s 
  \label{JNC}\;,
 \end{eqnarray}
 \label{EWQuarkCurrent}
\end{subequations}
where $g_{\rm v}^{\rm f}$ is the weak-vector charge of 
quark ``${\rm f}$'' expressed in terms of its weak isospin
and the weak mixing angle. That is, 
\begin{equation} 
 g_{\rm v}^{\rm f} = 
 2T_{3}^{\rm f}-4 Q_{\rm f}\sin^{2}\theta_{\rm w} =
 \begin{cases} 
  +1-\frac{8}{3}\sin^{2}\theta_{\rm w} \simeq +0.384 
  & \text{if f=\{u,c,t\},} \\
  -1+\frac{4}{3}\sin^{2}\theta_{\rm w} \simeq -0.692 
  & \text{if f=\{d,s,b\}.} \\
 \end{cases}
\end{equation}
Note that we are assuming that the heavy-quark ({c,b,t}) content of
the nucleon is negligible. Also note that the weak-vector charge of
the quarks are to a good approximation equal to the negative of the
electromagnetic charge of its weak isospin partner. This is the main
reason why the neutral $Z^{0}$-boson is an excellent probe of the
neutron density. Assuming isospin invariance, namely, that the 
up(down)-quark distribution in the proton equals the down(up)-quark 
distribution in the neutron and that the strange-quark distribution is
equal in both, the proton and neutron electromagnetic currents may 
be written in the following way:
\begin{subequations}    
 \begin{eqnarray}
  J^{\mu}_{\rm EM}(p) \equiv \langle p|\hat{J}^{\mu}_{\rm EM}|p\rangle
  &\!=\!&
   \frac{2}{3} V^{\mu}_{\rm \,u} -
  \frac{1}{3}  V^{\mu}_{\rm \,d}-
  \frac{1}{3}  V^{\mu}_{\rm \,s} \,,\\
  J^{\mu}_{\rm EM}(n) \equiv \langle n|\hat{J}^{\mu}_{\rm EM}|n\rangle
  &\!=\!&
   \frac{2}{3}V^{\mu}_{\rm \,d}  -
  \frac{1}{3} V^{\mu}_{\rm \,u} -
  \frac{1}{3} V^{\mu}_{\rm \,s} \,,
 \end{eqnarray}
 \label{EMQuarkCurrent}
\end{subequations}
where $V^{\mu}_{\rm \,u}$, $V^{\mu}_{\rm \,d}$, and $V^{\mu}_{\rm \,s}$
are matrix elements of the respective quark vector currents in 
the proton. From the above equations one can determine 
$V^{\mu}_{\rm \,u}$ and $V^{\mu}_{\rm \,d}$ in terms of the two
electromagnetic currents and $V^{\mu}_{\rm \,s}$. That is,
\begin{subequations}    
 \begin{eqnarray}
  V^{\mu}_{\rm \,u} &\!=\!& 
  2J^{\mu}_{\rm EM}(p) +  J^{\mu}_{\rm EM}(n) + V^{\mu}_{\rm \,s}
  \,, \\
  V^{\mu}_{\rm \,d} &\!=\!& 
  J^{\mu}_{\rm EM}(p) + 2 J^{\mu}_{\rm EM}(n) + V^{\mu}_{\rm \,s}
  \,.
 \end{eqnarray}
 \label{VuVd}
\end{subequations}
In turn, these relations may be used to express the matrix elements of 
the weak neutral current in terms of the corresponding matrix elements 
of the electromagnetic current plus the strange-quark contribution.
Inserting these relations into Eq.\,(\ref{JNC}) one obtains:
\begin{subequations}    
 \begin{eqnarray}
  J^{\mu}_{\rm NC}(p) &\!=\!& g_{\rm v}^{\rm p} J^{\mu}_{\rm EM}(p) +  
  g_{\rm v}^{\rm n} J^{\mu}_{\rm EM}(n) + 
  \xi_{\rm v}^{\rm (0)} V^{\mu}_{\rm \,s} \,,\\
  J^{\mu}_{\rm NC}(n) &\!=\!& g_{\rm v}^{\rm n} J^{\mu}_{\rm EM}(p) +  
  g_{\rm v}^{\rm p} J^{\mu}_{\rm EM}(n) + 
  \xi_{\rm v}^{\rm (0)} V^{\mu}_{\rm \,s} \,,
 \end{eqnarray}
 \label{NCQuarkCurrent}
\end{subequations}
where proton, neutron, and {\sl singlet} weak-vector charges are
given---including radiative corrections---by\,\cite{Liu:2007yi}, 
\begin{subequations}    
 \begin{eqnarray}
  g_{\rm v}^{\rm p} &\!=\!&  2g_{\rm v}^{\rm u} + g_{\rm v}^{\rm d} = 
 (1-4\sin^{2}\theta_{\rm w})(1+R_{\rm v}^{\rm p}) \approx 0.0712 \,, \\
  g_{\rm v}^{\rm n} &\!=\!&  g_{\rm v}^{\rm u} + 2g_{\rm v}^{\rm d} = 
 -(1+R_{\rm v}^{\rm n}) \approx -0.9877 \,, \\
  \xi_{\rm v}^{(0)} &\!=\!&  
  g_{\rm v}^{\rm u} + g_{\rm v}^{\rm d} + g_{\rm v}^{\rm s} =  
  -(1+R_{\rm v}^{(0)}) \approx -0.9877 \,.
 \end{eqnarray}
 \label{WeakCharges}
\end{subequations}
Note that these values are very close to the ones used in 
Ref.\,\cite{Horowitz:2012tj} which are the ones that have 
been adopted here ({\sl i.e.,} $g_{\rm v}^{\rm p}\!=\!0.0721$ 
and $g_{\rm v}^{\rm n}\!=\!-0.9878$). Finally, electric and 
magnetic Sachs form factors for the weak neutral current 
may be expressed in terms of the corresponding electromagnetic 
form factors plus the strange-quark contribution as follows:
\begin{subequations}    
 \begin{eqnarray}
   \widetilde{G}_{\rm E,M}^{\,p}(Q^{2}) &\!=\!&
   g_{\rm v}^{\rm p}G_{\rm E,M}^{\,p}(Q^{2})  +
   g_{\rm v}^{\rm n}G_{\rm E,M}^{\,n}(Q^{2})  +
   \xi_{\rm v}^{(0)}G_{\rm E,M}^{\,s}(Q^{2}) \,, \\
   \widetilde{G}_{\rm E,M}^{\,n}(Q^{2}) &\!=\!&
   g_{\rm v}^{\rm n}G_{\rm E,M}^{\,p}(Q^{2})  +
   g_{\rm v}^{\rm p}G_{\rm E,M}^{\,n}(Q^{2})  +
   \xi_{\rm v}^{(0)}G_{\rm E,M}^{\,s}(Q^{2}) \,.
 \end{eqnarray}
 \label{WeakFFs}
\end{subequations}
In particular, using these relations we obtain the following
expressions for the nucleon weak-charge radii:
\begin{subequations}    
 \begin{eqnarray}
   g_{\rm v}^{\rm p}\tilde{r}_{p}^{2} &\!=\!&
   g_{\rm v}^{\rm p}r_{p}^{2} +
   g_{\rm v}^{\rm n}r_{n}^{2} +
   \xi_{\rm v}^{(0)}r_{s}^{2} \,, \\
   g_{\rm v}^{\rm n}\tilde{r}_{n}^{2} &\!=\!&
   g_{\rm v}^{\rm n}r_{p}^{2} +
   g_{\rm v}^{\rm p}r_{n}^{2} +
   \xi_{\rm v}^{(0)}r_{s}^{2} \,.
 \end{eqnarray}
 \label{Weakr2s}
\end{subequations}
Similar expressions follow in the case of the nucleon weak 
magnetic moments:
\begin{subequations}    
 \begin{eqnarray}
   \tilde{\mu}_{p} &\!=\!&
    g_{\rm v}^{\rm p}\mu_{p} +
    g_{\rm v}^{\rm n}\mu_{n} +
    \xi_{\rm v}^{(0)}\mu_{s} \,, \\
  \tilde{\mu}_{n} &\!=\!&
   g_{\rm v}^{\rm n}\mu_{p} +
   g_{\rm v}^{\rm p}\mu_{n} +
   \xi_{\rm v}^{(0)}\mu_{s} \,.
 \end{eqnarray}
 \label{Weakmus}
\end{subequations}
Note that the mean-square strange radius and strange magnetic 
moment are defined as
\begin{equation}  
  r_{s}^{2}\!=\!-6
 \frac{dG_{\rm E}^{\,s}(Q^{2})}{dQ^{2}}\Big|_{\!Q^{2}\!=\!0}
 \quad {\rm and} \quad
 \mu_{s}\!=\!G_{\rm M}^{\,s}(Q^{2}\!=\!0)\,.
\end{equation}  

\bibliography{../ReferencesJP}

\begin{thebibliography}{32}
\expandafter\ifx\csname natexlab\endcsname\relax\def\natexlab#1{#1}\fi
\expandafter\ifx\csname bibnamefont\endcsname\relax
  \def\bibnamefont#1{#1}\fi
\expandafter\ifx\csname bibfnamefont\endcsname\relax
  \def\bibfnamefont#1{#1}\fi
\expandafter\ifx\csname citenamefont\endcsname\relax
  \def\citenamefont#1{#1}\fi
\expandafter\ifx\csname url\endcsname\relax
  \def\url#1{\texttt{#1}}\fi
\expandafter\ifx\csname urlprefix\endcsname\relax\def\urlprefix{URL }\fi
\providecommand{\bibinfo}[2]{#2}
\providecommand{\eprint}[2][]{\url{#2}}

\bibitem[{\citenamefont{Abrahamyan et~al.}(2012)\citenamefont{Abrahamyan,
  Ahmed, Albataineh, Aniol, Armstrong et~al.}}]{Abrahamyan:2012gp}
\bibinfo{author}{\bibfnamefont{S.}~\bibnamefont{Abrahamyan}},
  \bibinfo{author}{\bibfnamefont{Z.}~\bibnamefont{Ahmed}},
  \bibinfo{author}{\bibfnamefont{H.}~\bibnamefont{Albataineh}},
  \bibinfo{author}{\bibfnamefont{K.}~\bibnamefont{Aniol}},
  \bibinfo{author}{\bibfnamefont{D.}~\bibnamefont{Armstrong}},
  \bibnamefont{et~al.}, \bibinfo{journal}{Phys. Rev. Lett.}
  \textbf{\bibinfo{volume}{108}}, \bibinfo{pages}{112502}
  (\bibinfo{year}{2012}).

\bibitem[{\citenamefont{Horowitz et~al.}(2012)\citenamefont{Horowitz, Ahmed,
  Jen, Rakhman, Souder et~al.}}]{Horowitz:2012tj}
\bibinfo{author}{\bibfnamefont{C. J.}~\bibnamefont{Horowitz}},
  \bibinfo{author}{\bibfnamefont{Z.}~\bibnamefont{Ahmed}},
  \bibinfo{author}{\bibfnamefont{C.}~\bibnamefont{Jen}},
  \bibinfo{author}{\bibfnamefont{A.}~\bibnamefont{Rakhman}},
  \bibinfo{author}{\bibfnamefont{P.}~\bibnamefont{Souder}},
  \bibnamefont{et~al.}, \bibinfo{journal}{Phys. Rev.}
  \textbf{\bibinfo{volume}{C85}}, \bibinfo{pages}{032501}
  (\bibinfo{year}{2012}).

\bibitem[{\citenamefont{Horowitz}(1998)}]{Horowitz:1998vv}
\bibinfo{author}{\bibfnamefont{C. J.}~\bibnamefont{Horowitz}},
  \bibinfo{journal}{Phys. Rev.} \textbf{\bibinfo{volume}{C57}},
  \bibinfo{pages}{3430} (\bibinfo{year}{1998}).

\bibitem[{\citenamefont{Roca-Maza et~al.}(2011)\citenamefont{Roca-Maza,
  Centelles, Vinas, and Warda}}]{RocaMaza:2011pm}
\bibinfo{author}{\bibfnamefont{X.}~\bibnamefont{Roca-Maza}},
  \bibinfo{author}{\bibfnamefont{M.}~\bibnamefont{Centelles}},
  \bibinfo{author}{\bibfnamefont{X.}~\bibnamefont{Vi\~nas}}, \bibnamefont{and}
  \bibinfo{author}{\bibfnamefont{M.}~\bibnamefont{Warda}},
  \bibinfo{journal}{Phys. Rev. Lett.} \textbf{\bibinfo{volume}{106}},
  \bibinfo{pages}{252501} (\bibinfo{year}{2011}).

\bibitem[{\citenamefont{Horowitz et~al.}(2001)\citenamefont{Horowitz, Pollock,
  Souder, and Michaels}}]{Horowitz:1999fk}
\bibinfo{author}{\bibfnamefont{C.~J.} \bibnamefont{Horowitz}},
  \bibinfo{author}{\bibfnamefont{S.~J.} \bibnamefont{Pollock}},
  \bibinfo{author}{\bibfnamefont{P.~A.} \bibnamefont{Souder}},
  \bibnamefont{and} \bibinfo{author}{\bibfnamefont{R.}~\bibnamefont{Michaels}},
  \bibinfo{journal}{Phys. Rev.} \textbf{\bibinfo{volume}{C63}},
  \bibinfo{pages}{025501} (\bibinfo{year}{2001}).

\bibitem[{\citenamefont{Angeli}(2004)}]{Angeli:2004}
\bibinfo{author}{\bibfnamefont{I.}~\bibnamefont{Angeli}}, \bibinfo{journal}{At.
  Data Nucl. Data Tables} \textbf{\bibinfo{volume}{87}}, \bibinfo{pages}{185}
  (\bibinfo{year}{2004}).

\bibitem[{\citenamefont{Paschke et~al.}(2012)\citenamefont{Paschke, Kumar,
  Michaels, Souder, and Urciuoli}}]{PREXII:2012}
\bibinfo{author}{\bibfnamefont{K.}~\bibnamefont{Paschke}},
  \bibinfo{author}{\bibfnamefont{K.}~\bibnamefont{Kumar}},
  \bibinfo{author}{\bibfnamefont{R.}~\bibnamefont{Michaels}},
  \bibinfo{author}{\bibfnamefont{P.~A.} \bibnamefont{Souder}},
  \bibnamefont{and} \bibinfo{author}{\bibfnamefont{G.~M.}
  \bibnamefont{Urciuoli}} (\bibinfo{year}{2012}),
  \urlprefix\url{http://hallaweb.jlab.org/parity/prex/prexII.pdf}.

\bibitem[{\citenamefont{Mammei et~al.}(2012)\citenamefont{Mammei, Michaels,
  Paschke, Riordan, and Souder}}]{CREX:2012}
\bibinfo{author}{\bibfnamefont{J.}~\bibnamefont{Mammei}},
  \bibinfo{author}{\bibfnamefont{R.}~\bibnamefont{Michaels}},
  \bibinfo{author}{\bibfnamefont{K.}~\bibnamefont{Paschke}},
  \bibinfo{author}{\bibfnamefont{S.}~\bibnamefont{Riordan}}, \bibnamefont{and}
  \bibinfo{author}{\bibfnamefont{P.~A.} \bibnamefont{Souder}}
  (\bibinfo{year}{2012}),
  \urlprefix\url{http://hallaweb.jlab.org/parity/prex/c-rex/c-rex.pdf}.

\bibitem[{\citenamefont{Piekarewicz et~al.}(2012)\citenamefont{Piekarewicz,
  Agrawal, Col\`o, Nazarewicz, Paar et~al.}}]{Piekarewicz:2012pp}
\bibinfo{author}{\bibfnamefont{J.}~\bibnamefont{Piekarewicz}},
  \bibinfo{author}{\bibfnamefont{B.}~\bibnamefont{Agrawal}},
  \bibinfo{author}{\bibfnamefont{G.}~\bibnamefont{Col\`o}},
  \bibinfo{author}{\bibfnamefont{W.}~\bibnamefont{Nazarewicz}},
  \bibinfo{author}{\bibfnamefont{N.}~\bibnamefont{Paar}}, \bibnamefont{et~al.}
  \bibinfo{journal}{Phys. Rev.} \textbf{\bibinfo{volume}{C85}},
  \bibinfo{pages}{041302(R)} (\bibinfo{year}{2012}).

\bibitem[{\citenamefont{Pollock et~al.}(1992)\citenamefont{Pollock, Fortson,
  and Wilets}}]{Pollock:1992mv}
\bibinfo{author}{\bibfnamefont{S.~J.} \bibnamefont{Pollock}},
  \bibinfo{author}{\bibfnamefont{E.~N.} \bibnamefont{Fortson}},
  \bibnamefont{and} \bibinfo{author}{\bibfnamefont{L.}~\bibnamefont{Wilets}},
  \bibinfo{journal}{Phys. Rev.} \textbf{\bibinfo{volume}{C46}},
  \bibinfo{pages}{2587} (\bibinfo{year}{1992}).

\bibitem[{\citenamefont{Sil et~al.}(2005)\citenamefont{Sil, Centelles, Vinas,
  and Piekarewicz}}]{Sil:2005tg}
\bibinfo{author}{\bibfnamefont{T.}~\bibnamefont{Sil}},
  \bibinfo{author}{\bibfnamefont{M.}~\bibnamefont{Centelles}},
  \bibinfo{author}{\bibfnamefont{X.}~\bibnamefont{Vi\~nas}}, \bibnamefont{and}
  \bibinfo{author}{\bibfnamefont{J.}~\bibnamefont{Piekarewicz}},
  \bibinfo{journal}{Phys. Rev.} \textbf{\bibinfo{volume}{C71}},
  \bibinfo{pages}{045502} (\bibinfo{year}{2005}).

\bibitem[{\citenamefont{Tsang et~al.}(2004)}]{Tsang:2004zz}
\bibinfo{author}{\bibfnamefont{M.~B.} \bibnamefont{Tsang}}
  \bibnamefont{et~al.}, \bibinfo{journal}{Phys. Rev. Lett.}
  \textbf{\bibinfo{volume}{92}}, \bibinfo{pages}{062701}
  (\bibinfo{year}{2004}).

\bibitem[{\citenamefont{Chen et~al.}(2005)\citenamefont{Chen, Ko, and
  Li}}]{Chen:2004si}
\bibinfo{author}{\bibfnamefont{L.-W.} \bibnamefont{Chen}},
  \bibinfo{author}{\bibfnamefont{C.~M.} \bibnamefont{Ko}}, \bibnamefont{and}
  \bibinfo{author}{\bibfnamefont{B.-A.} \bibnamefont{Li}},
  \bibinfo{journal}{Phys. Rev. Lett.} \textbf{\bibinfo{volume}{94}},
  \bibinfo{pages}{032701} (\bibinfo{year}{2005}).

\bibitem[{\citenamefont{Steiner and Li}(2005)}]{Steiner:2005rd}
\bibinfo{author}{\bibfnamefont{A.~W.} \bibnamefont{Steiner}} \bibnamefont{and}
  \bibinfo{author}{\bibfnamefont{B.-A.} \bibnamefont{Li}},
  \bibinfo{journal}{Phys. Rev.} \textbf{\bibinfo{volume}{C72}},
  \bibinfo{pages}{041601} (\bibinfo{year}{2005}).

\bibitem[{\citenamefont{Shetty et~al.}(2007)\citenamefont{Shetty, Yennello, and
  Souliotis}}]{Shetty:2007zg}
\bibinfo{author}{\bibfnamefont{D.~V.} \bibnamefont{Shetty}},
  \bibinfo{author}{\bibfnamefont{S.~J.} \bibnamefont{Yennello}},
  \bibnamefont{and} \bibinfo{author}{\bibfnamefont{G.~A.}
  \bibnamefont{Souliotis}}, \bibinfo{journal}{Phys. Rev.}
  \textbf{\bibinfo{volume}{C76}}, \bibinfo{pages}{024606}
  (\bibinfo{year}{2007}).

\bibitem[{\citenamefont{Tsang et~al.}(2009)}]{Tsang:2008fd}
\bibinfo{author}{\bibfnamefont{M.~B.} \bibnamefont{Tsang}}
  \bibnamefont{et~al.}, \bibinfo{journal}{Phys. Rev. Lett.}
  \textbf{\bibinfo{volume}{102}}, \bibinfo{pages}{122701}
  (\bibinfo{year}{2009}).

\bibitem[{\citenamefont{Horowitz and
  Piekarewicz}(2001{\natexlab{a}})}]{Horowitz:2000xj}
\bibinfo{author}{\bibfnamefont{C.~J.} \bibnamefont{Horowitz}} \bibnamefont{and}
  \bibinfo{author}{\bibfnamefont{J.}~\bibnamefont{Piekarewicz}},
  \bibinfo{journal}{Phys. Rev. Lett.} \textbf{\bibinfo{volume}{86}},
  \bibinfo{pages}{5647} (\bibinfo{year}{2001}{\natexlab{a}}).

\bibitem[{\citenamefont{Horowitz and
  Piekarewicz}(2001{\natexlab{b}})}]{Horowitz:2001ya}
\bibinfo{author}{\bibfnamefont{C.~J.} \bibnamefont{Horowitz}} \bibnamefont{and}
  \bibinfo{author}{\bibfnamefont{J.}~\bibnamefont{Piekarewicz}},
  \bibinfo{journal}{Phys. Rev.} \textbf{\bibinfo{volume}{C64}},
  \bibinfo{pages}{062802} (\bibinfo{year}{2001}{\natexlab{b}}).

\bibitem[{\citenamefont{Horowitz and Piekarewicz}(2002)}]{Horowitz:2002mb}
\bibinfo{author}{\bibfnamefont{C.~J.} \bibnamefont{Horowitz}} \bibnamefont{and}
  \bibinfo{author}{\bibfnamefont{J.}~\bibnamefont{Piekarewicz}},
  \bibinfo{journal}{Phys. Rev.} \textbf{\bibinfo{volume}{C66}},
  \bibinfo{pages}{055803} (\bibinfo{year}{2002}).

\bibitem[{\citenamefont{Carriere et~al.}(2003)\citenamefont{Carriere, Horowitz,
  and Piekarewicz}}]{Carriere:2002bx}
\bibinfo{author}{\bibfnamefont{J.}~\bibnamefont{Carriere}},
  \bibinfo{author}{\bibfnamefont{C.~J.} \bibnamefont{Horowitz}},
  \bibnamefont{and}
  \bibinfo{author}{\bibfnamefont{J.}~\bibnamefont{Piekarewicz}},
  \bibinfo{journal}{Astrophys. J.} \textbf{\bibinfo{volume}{593}},
  \bibinfo{pages}{463} (\bibinfo{year}{2003}).

\bibitem[{\citenamefont{Steiner et~al.}(2005)\citenamefont{Steiner, Prakash,
  Lattimer, and Ellis}}]{Steiner:2004fi}
\bibinfo{author}{\bibfnamefont{A.~W.} \bibnamefont{Steiner}},
  \bibinfo{author}{\bibfnamefont{M.}~\bibnamefont{Prakash}},
  \bibinfo{author}{\bibfnamefont{J.~M.} \bibnamefont{Lattimer}},
  \bibnamefont{and} \bibinfo{author}{\bibfnamefont{P.~J.} \bibnamefont{Ellis}},
  \bibinfo{journal}{Phys. Rept.} \textbf{\bibinfo{volume}{411}},
  \bibinfo{pages}{325} (\bibinfo{year}{2005}).

\bibitem[{\citenamefont{Li and Steiner}(2006)}]{Li:2005sr}
\bibinfo{author}{\bibfnamefont{B.-A.} \bibnamefont{Li}} \bibnamefont{and}
  \bibinfo{author}{\bibfnamefont{A.~W.} \bibnamefont{Steiner}},
  \bibinfo{journal}{Phys. Lett.} \textbf{\bibinfo{volume}{B642}},
  \bibinfo{pages}{436} (\bibinfo{year}{2006}).

\bibitem[{\citenamefont{Ito and Gross}(1993)}]{Ito:1993au}
\bibinfo{author}{\bibfnamefont{H.}~\bibnamefont{Ito}} \bibnamefont{and}
  \bibinfo{author}{\bibfnamefont{F.}~\bibnamefont{Gross}},
  \bibinfo{journal}{Phys. Rev. Lett.} \textbf{\bibinfo{volume}{71}},
  \bibinfo{pages}{2555} (\bibinfo{year}{1993}).

\bibitem[{\citenamefont{Ong et~al.}(2010)\citenamefont{Ong, Berengut, and
  Flambaum}}]{Ong:2010gf}
\bibinfo{author}{\bibfnamefont{A.}~\bibnamefont{Ong}},
  \bibinfo{author}{\bibfnamefont{J.}~\bibnamefont{Berengut}}, \bibnamefont{and}
  \bibinfo{author}{\bibfnamefont{V.}~\bibnamefont{Flambaum}},
  \bibinfo{journal}{Phys. Rev.} \textbf{\bibinfo{volume}{C82}},
  \bibinfo{pages}{014320} (\bibinfo{year}{2010}).

\bibitem[{\citenamefont{Musolf et~al.}(1994)\citenamefont{Musolf, Donnelly,
  Dubach, Pollock, Kowalski et~al.}}]{Musolf:1993tb}
\bibinfo{author}{\bibfnamefont{M.}~\bibnamefont{Musolf}},
  \bibinfo{author}{\bibfnamefont{T.}~\bibnamefont{Donnelly}},
  \bibinfo{author}{\bibfnamefont{J.}~\bibnamefont{Dubach}},
  \bibinfo{author}{\bibfnamefont{S.}~\bibnamefont{Pollock}},
  \bibinfo{author}{\bibfnamefont{S.}~\bibnamefont{Kowalski}},
  \bibnamefont{et~al.}, \bibinfo{journal}{Phys. Rept.}
  \textbf{\bibinfo{volume}{239}}, \bibinfo{pages}{1} (\bibinfo{year}{1994}).

\bibitem[{\citenamefont{Serot and Walecka}(1986)}]{Serot:1984ey}
\bibinfo{author}{\bibfnamefont{B.~D.} \bibnamefont{Serot}} \bibnamefont{and}
  \bibinfo{author}{\bibfnamefont{J.~D.} \bibnamefont{Walecka}},
  \bibinfo{journal}{Adv. Nucl. Phys.} \textbf{\bibinfo{volume}{16}},
  \bibinfo{pages}{1} (\bibinfo{year}{1986}).

\bibitem[{\citenamefont{Todd-Rutel and Piekarewicz}(2005)}]{Todd-Rutel:2005fa}
\bibinfo{author}{\bibfnamefont{B.~G.} \bibnamefont{Todd-Rutel}}
  \bibnamefont{and}
  \bibinfo{author}{\bibfnamefont{J.}~\bibnamefont{Piekarewicz}},
  \bibinfo{journal}{Phys. Rev. Lett} \textbf{\bibinfo{volume}{95}},
  \bibinfo{pages}{122501} (\bibinfo{year}{2005}).

\bibitem[{\citenamefont{Kelly}(2002)}]{Kelly:2002if}
\bibinfo{author}{\bibfnamefont{J.~J.} \bibnamefont{Kelly}},
  \bibinfo{journal}{Phys. Rev.} \textbf{\bibinfo{volume}{C66}},
  \bibinfo{pages}{065203} (\bibinfo{year}{2002}).

\bibitem[{\citenamefont{Nakamura et~al.}(2010)}]{Nakamura:2010zzi}
\bibinfo{author}{\bibfnamefont{K.}~\bibnamefont{Nakamura}} \bibnamefont{et~al.}
  (\bibinfo{collaboration}{Particle Data Group}), \bibinfo{journal}{J. Phys. G}
  \textbf{\bibinfo{volume}{G37}}, \bibinfo{pages}{075021}
  (\bibinfo{year}{2010}).

\bibitem[{\citenamefont{Lalazissis et~al.}(1997)\citenamefont{Lalazissis,
  Konig, and Ring}}]{Lalazissis:1996rd}
\bibinfo{author}{\bibfnamefont{G.~A.} \bibnamefont{Lalazissis}},
  \bibinfo{author}{\bibfnamefont{J.}~\bibnamefont{Konig}}, \bibnamefont{and}
  \bibinfo{author}{\bibfnamefont{P.}~\bibnamefont{Ring}},
  \bibinfo{journal}{Phys. Rev.} \textbf{\bibinfo{volume}{C55}},
  \bibinfo{pages}{540} (\bibinfo{year}{1997}).

\bibitem[{\citenamefont{Lalazissis et~al.}(1999)\citenamefont{Lalazissis,
  Raman, and Ring}}]{Lalazissis:1999}
\bibinfo{author}{\bibfnamefont{G.~A.} \bibnamefont{Lalazissis}},
  \bibinfo{author}{\bibfnamefont{S.}~\bibnamefont{Raman}}, \bibnamefont{and}
  \bibinfo{author}{\bibfnamefont{P.}~\bibnamefont{Ring}}, \bibinfo{journal}{At.
  Data Nucl. Data Tables} \textbf{\bibinfo{volume}{71}}, \bibinfo{pages}{1}
  (\bibinfo{year}{1999}).

\bibitem[{\citenamefont{Liu et~al.}(2007)\citenamefont{Liu, McKeown, and
  Ramsey-Musolf}}]{Liu:2007yi}
\bibinfo{author}{\bibfnamefont{J.}~\bibnamefont{Liu}},
  \bibinfo{author}{\bibfnamefont{R.~D.} \bibnamefont{McKeown}},
  \bibnamefont{and} \bibinfo{author}{\bibfnamefont{M.~J.}
  \bibnamefont{Ramsey-Musolf}}, \bibinfo{journal}{Phys. Rev.}
  \textbf{\bibinfo{volume}{C76}}, \bibinfo{pages}{025202}
  (\bibinfo{year}{2007}).

\end{thebibliography}
\vfill\eject
\end{document}